\documentclass{article}

\usepackage{amsmath,amssymb,epsf,epsfig,a4wide}

\begin{document}
\begin{titlepage}

\begin{centering}
\begin{flushright}
hep-th/0106146
\end{flushright}

\vspace{0.1in}

{\Large {\bf Impulse Action on $D$-particles 
in Robertson-Walker Space Times,
Higher-Order Logarithmic Conformal Algebras and Cosmological Horizons}}

\vspace{0.4in}

{\bf Elias Gravanis and Nick E. Mavromatos } \\
\vspace{0.2in}
Department of Physics, Theoretical Physics, King's College London,\\
Strand, London WC2R 2LS, United Kingdom.

\vspace{0.4in}
 {\bf Abstract}

\end{centering}

\vspace{0.2in}

{\small We demonstrate that an impulse action (`recoil') 
on a D-particle embedded in a
(four-dimensional) cosmological Robertson-Walker (RW) spacetime
is described, in a $\sigma$-model 
framework, 
by a suitably extended higher-order logarithmic world-sheet
algebra of relevant deformations. 
We study in some detail the algebra of the appropriate
two-point correlators, and give  
a careful discussion as to how one can approach the world-sheet
renormalization group infrared fixed point, in the neighborhood of
which the
logarithmic algebra is valid.
It is found that, if 
the initial RW spacetime does not have 
cosmological 
horizons, then there is no problem in approaching the fixed point. 
However, in the presence of horizons, there are world-sheet divergences
which imply  
the need for Liouville dressing in order to approach 
the fixed point in the correct way. 
A detailed analysis on the subtle subtraction of these divergences
in the latter case is given.  
In both cases, 
at the fixed point, the recoil-induced spacetime 
is nothing other than
a coordinate transformation of the initial spacetime into the 
rest frame of the recoiling D-particle. 
However, in the horizon case, if one identifies 
the Liouville mode with the target time, 
which 
expresses physically the back reaction of the recoiling D-particle 
onto the spacetime structure,  
it is found that the induced spacetime
distortion results in the removal of the initial cosmological horizon
and the eventual stopping of the acceleration of the Universe. 
In this latter sense, 
our model may be thought of as a conformal field theory description  
of a (toy) Universe characterized by a sort of `phase transition'
at the moment of impulse, implying a time-varying 
speed of light.}

\vspace{0.8in}
\begin{flushleft}
June 2001 
\end{flushleft}

\end{titlepage}

\section{Introduction and Summary}

Placing D-branes in curved space times  
is not understood well at 
present. The main problem originates from the lack of knowledge of the 
complete dynamics of such solitonic objects. One would hope that 
such a knowledge 
would allow a proper study of the back reaction of such objects
onto the surrounding space time geometry (distortion), and eventually 
a consistent discussion of their dynamics in curved spacetimes.  
Some modest steps towards an incorporation of curved space time
effects in D-brane dynamics have been taken in the recent literature
from a number of authors~\cite{curved}. 
These works are dealing directly 
with world volume effects of D-branes and in some cases 
string dualities are used in order to discuss the effects of space time 
curvature. 

A different approach
has been adopted in~\cite{kogan,kmw,szabo,recoil}, in which 
we have attempted to approach some aspects of the problem from a world 
sheet view point, which is probably suitable for a study of
the effects of 
the (string) excitations of the heavy brane. 
We have concentrated mainly on heavy $D$-particles,
embedded in a {\it flat} target background space time.
We have discussed the instantaneous action (impulse) 
of a `force' on a heavy $D$-particle. The impulse 
may be viewed either as 
a consequence of 
`trapping' of a {\it macrosopic number} of   
closed string states on the defect,
and their eventual splitting into pairs of open strings, 
or, in a different context,
as the result of a more general phenomenon associated 
with the {\it sudden} appearance of such defects. 
Our world sheet approach 
is a valid approximation only if one looks at times 
{\it long after} the event. 
Such impulse approximations usually characterize classical 
phenomena. In our picture we view the whole 
process as a {\it semi-classical} phenomenon, due to the fact that 
the process involves open string {\it recoil} 
excitations of the heavy $D$-particle, which are {\it quantum} in nature.
It is this point of view that we shall adopt in the present article.  

Such an approach should be distinguished 
from the 
problem of studying single-string 
scattering of a $D$-particle with closed string states
in flat space times~\cite{paban}.
We have shown in \cite{kogan,kmw,szabo,recoil} that 
for a $D$-particle embedded 
in a $d$-dimensional {\it flat Minkowski} space time
such an impulse action 
is described by a world-sheet $\sigma$-model deformed by 
appropriate `recoil' operators, which obey a logarithmic conformal 
algebra~\cite{lcft}. The appearance of such algebras, which lie 
on the border line between conformal field theories 
and general renormalizable 
field theories in the 
two-dimensional world sheet, but can still be classified by conformal 
data, is associated with the fact that an impulse action (recoil) 
describes a {\it change} 
of the string/D-particle background, and as such it cannot
be described by conformal symmetry all along.   
The {\it transition} between the two asymptotic states of the system 
before and (long) after the event 
is precisely described by deforming the associated $\sigma$-model
by operators which {\it spoil} the conformal symmetry.

Indeed, 
the recoil operators are {\it relevant} from a 
world-sheet renormalization-group view point~\cite{kmw}, and thus the 
induced string theory becomes non-critical, in need of Liouville 
dressing~\cite{ddk} 
in order to restore the conformal symmetry. 
The dressing results in the appearance of target-space metric
distortion~\cite{recoil},
which - under the identification of the Liouville mode with 
the time~\cite{emn} -
is interpreted as a backreaction of the recoiling $D$-particle
defect onto the surrounding (initially flat) space time. 
Under such an impulse/recoil, there is in general an induced vacuum energy,
which can even become time dependent~\cite{emncosmo}.
Such time dependent vacuum energies in Cosmology have recently attracted
a lot of attention as a challenge for string theory~\cite{challenge},
given that in certain cases the corresponding Universes 
are characterized by cosmological horizons, and hence  
a field-theoretic $S$ matrix cannot be defined for asymptotic states.
From the point of view of Liouville string
such a situation is expected~\cite{emnsmatrix}, due to the fact that 
in Liouville strings, with the time identified with the Liouville 
mode~\cite{emn}, a scattering matrix cannot be defined. 

In this work we shall attempt to extend the flat space time results 
of \cite{kogan,kmw,recoil} to the physically relevant 
case of a Robertson-Walker (RW) cosmological background space time.
Although, 
our results do not depend on the target space dimension, 
however, for definiteness we shall concentrate on the case of a
$D$-particle embedded in a four-dimensional RW spacetime. 
It must be stressed that we shall not attempt here to 
present a complete 
discussion of the associated 
space time curvature effects, which - as mentioned earlier - 
is a very difficult 
task, still unresolved.
Nevertheless, by concentrating on times 
much larger than the moment of impulse 
on the $D$-particle defect, one 
may 
ignore such effects to a satisfactory approximation.
As we shall see,  
our analysis    
produces results which look reasonable 
and are of sufficient  
interest 
to 
initiate further research.  

The vertex operators which describe the impulse 
in curved RW backgrounds 
obey a suitably extended (higher-order) logarithmic algebra.
The algebra is valid at, and in the neighborhood of, a non-trivial 
infrared fixed point of the world-sheet Renormalization Group.  
For a RW spacetime of scale factor of the form $t^p$, where $t$ 
is the target time, and $p > 1$ in the horizon case, the algebra 
is actually a set of logarithmic algebras up to order $[2p]$, 
which are classified by the appropriate 
higher-order Jordan 
blocks~\cite{lcft}.   

As in the flat case, 
which is obtained as a special limit of this
more general case, 
the recoil deformations are relevant operators 
from a world-sheet Renormalization-Group viewpoint.
One distinguishes two cases. In the first, the 
initial RW spacetime does not possess cosmological horizons.
In this case it is shown that the limit to the conformal world-sheet
non-trivial (infrared) fixed point can be taken smoothly without problems.
On the other hand, in the case where the initial spacetime has cosmological
horizons, such a limit is plagued by world-sheet divergences.   
These should 
be carefully subtracted in order
to allow for a smooth approach to the fixed point. 
A detailed discussion of how this can be done is presented.
In general, the divergences 
spoil the conformal invariance of the 
$\sigma$-model, thus implying the need  
for Liouville dressing~\cite{ddk} in order to properly
restore the conformal symmetry.

Moreover, a 
careful discussion of the matching between the results of the Liouville
dressing and those implied by the logarithmic algebra is given, 
which supports the possibility of identifying the world-sheet zero mode
of the Liouville field 
(viewed as a local renormalization-group scale on the world sheet) 
with the target time. One distinguishes various cases
which depend on whether the underlying theory lives in its critical 
dimension, and thus the only source of not criticality is 
the impulse action, 
or not. Such an identification  
induces target-space metric deformations, which are 
responsible  
for the {\it removal} of the cosmological horizon 
of the initial 
spacetime background, and the  
stopping of the acceleration of the Universe. 
Essentially the situation implies  
an effective time-dependent light velocity after the moment of impulse,
which is responsible for the removal of the cosmological horizon.
From this point of view our work may thus seem to provide a 
conformal-field-theory framework for a proper treatment 
of such time-varying speed of light
scenaria~\cite{moffat} in 
the context of non-critical string theory~\cite{emnsmatrix}.

\section{Recoiling D-particles in Robertson-Walker Backgrounds}

\subsection{Geodesic Paths and Recoil} 

Let us consider a $D$-particle, located (for convenience) at the origin 
of the spatial coordinates of a four-dimensional space time, which at 
a time $t_0$ experiences an impulse. 
In a $\sigma$-model framework, the trajectory of the $D$-particle 
$y^i(t)$, $i$ a spatial index,   
is described by inserting the following
vertex operator
\begin{equation} \label{path}
V = \int _{\partial \Sigma} G_{ij}y^j(t)\partial_n X^i 
\end{equation}
where $G_{ij}$ denotes the spatial components of the metric, 
$\partial \Sigma$ denotes the world-sheet boundary, 
$\partial _n$ is a normal world-sheet derivative, 
$X^i$ are $\sigma$-model fields obeying Dirichlet boundary conditions 
on the world sheet, 
and $t$ is a $\sigma$-model field obeying Neumann boundary conditions
on the world sheet, whose zero mode is the target time. 

This is the basic vertex deformation which we assume 
to describe the motion of a $D$-particle in a curved geometry
to leading order at least, where spacetime back reaction 
and curvature effects are assumed weak.
Such vertex deformations may be viewed as a 
generalization of the flat-target-space case~\cite{dparticle}.

Perhaps a formally more desirable approach 
towards the construction of the complete vertex operator 
would be to start from 
a T-dual (Neumann) picture, where the deformation (\ref{path}) 
should correspond to a proper Wilson loop operator of an
appropriate  
gauge vector field. Such loop operators are by construction independent
of the background geometry. One can 
then pass onto the Dirichlet picture by a T-duality transformation
viewed as a canonical transformation from a $\sigma$-model 
viewpoint~\cite{otto}. In principle, such a procedure  
would yield a complete form of the vertex operator in the Dirichlet 
picture, 
describing the path of a $D$-particle in a curved geometry.
Unfortunately, such a procedure is not free from ambiguities
at a quantum level~\cite{otto}, 
which are still unresolved for general curved backgrounds. 
Therefore, for our purposes here, we shall consider the 
problem of writing a complete form for the 
operator (\ref{path}) in a RW spacetime 
background 
in the Dirichlet picture as an open issue. Nevertheless, for RW backgrounds
at large times, ignoring curvature effects proves to be a satisfactory
approximation, and in such a case 
one may consider the vertex operator (\ref{path}) 
as a sufficient description for the physical vertex operator 
of a $D$-particle. 
As we shall show below, the results
of such analyses appear reasonable and interesting enough 
to encourage further
studies along this direction.   

For times long after the event,   
the trajectory $y^i(t)$ will be that of free motion
in the 
curved space time under consideration. In the flat space time 
case, this trajectory was a straight line~\cite{dparticle,kmw,szabo}, 
and in the more general 
case here it will be simply the associated {\it geodesic}. 
Let us now determine its form, which will be essential in what follows. 

The space time assumes the form:
\begin{equation}\label{rwmetric}
ds^2 = -dt^2 + a(t)^2 (dX^i)^2  
\end{equation}
where $a(t)$ is the RW scale factor. We shall work with 
expanding RW space times with scale factors 
\begin{equation}  
a(t) =a_0 t^p, \qquad p \in R^+ 
\end{equation} 
The geodesic equations in this case read:
\begin{eqnarray}
{\ddot t} + pt^{2p-1}({\dot y}^i)^2 &=& 0 \nonumber \\
{\ddot y} + 2\frac{p}{t}({\dot y}^i) {\dot t}&=& 0
\end{eqnarray}
where the dot denotes differentiation with respect to 
the proper time $\tau$ of the $D$-particle. 

With initial conditions $y^i(t_0)=0$, and $dy^i/dt (t_0) \equiv v^i$, 
one easily finds that, for long times $t \gg t_0$ after the event,  
the solution acquires the form: 
\begin{equation}\label{pathexpre} 
y^i(t) =\frac{v^i}{1-2p}\left(t^{1-2p}t_0^{2p} - t_0 \right) + 
{\cal O}(t^{1-4p}), \qquad t  \gg t_0 
\end{equation}

To leading order in $t$, therefore, the appropriate vertex operator
(\ref{path}), describing the recoil of the $D$-particle, 
is:
\begin{equation}\label{path2} 
V=\int _{\partial \Sigma} a_0^2 \frac{v^i}{1-2p}\Theta (t-t_0)\left(tt_0^{2p}-t_0t^{2p}\right)
\partial_n X^i 
\end{equation} 
where $\Theta (t-t_0)$ is the Heaviside step function,
expressing an instantaneous action ({\it impulse}) 
on the $D$-particle
at $t=t_0$~\cite{kmw,recoil}. 
As we shall see later on,  
such deformed $\sigma$-models
may be viewed as 
providing rather generic mathematical prototypes for models 
involving phase transitions at early stages of the Universe, 
leading effectively 
to time-varying speed of light. 
In the context of the present work, therefore, we shall be
rather vague as far as the precise physical significance of the 
operator (\ref{path2}) is concerned, and merely exploit the 
consequences
of such deformations for the expansion of the RW spacetime after time $t_0$, 
from both a mathematical and physical viewpoint. 

\begin{figure}[t] 
\begin{centering} 
\epsfig{figure=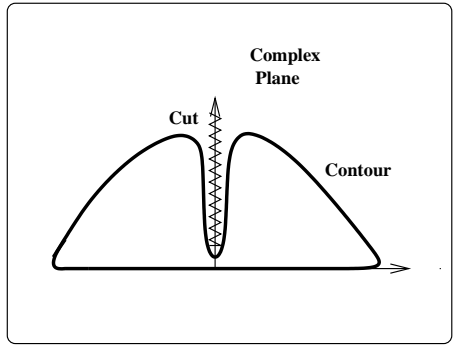, width=0.4\linewidth}
\caption{{\it Contour of integration in the complex plane to define 
the recoil operators ${\cal D}^{(q)}$, by proper treatment of the
associated cuts.}}
\end{centering} 
\label{cut}
\end{figure}

In \cite{kmw}, we have studied the case $p=0$, $a_0=1$, 
where the operators
assumed the form $t\Theta_\epsilon (t)$ to leading order in $t$,
where $\Theta_\epsilon (t)$ is the regulated form of the step
function, given by~\cite{kmw}:
\begin{equation}\label{rep} 
\Theta_\epsilon =-i\int_{-\infty}^{+\infty} 
\frac{d\omega}{2\pi} \frac{1}{\omega -i\epsilon}
e^{i\omega\,t}, \qquad \epsilon \rightarrow 0^+ 
\end{equation}
As discussed in that reference, this operator 
forms a logarithmic pair~\cite{lcft} with 
$\epsilon \Theta_\epsilon  (t)$, expressing physically fluctuations 
in the initial position of the $D$-particle.

In the current case, one may expand the integrand of (\ref{path2}) 
in a Taylor series in powers of $(t-t_0)$, which implies 
the presence of a series of operators, of the form 
$(t-t_0)^q\Theta_\epsilon (t-t_0)$, where $q $ takes on the values
$2p, 2p-1, \dots $, i.e. it is not an integer in general.  
In a direct generalization of the Fourier integral representation 
(\ref{rep}), we write in this case:
\begin{eqnarray}\label{opd}  
&~& {\cal D}^{(q)} 
\equiv v_i (t-t_0)^q \Theta_\epsilon (t-t_0)\partial_n X^i  
=v_i~N_q~\int _{-\infty}^{+\infty}d\omega 
\frac{1}{(\omega -i\epsilon)^{q+1}}~e^{i\omega(t-t_0)}\partial_nX^i~,
\nonumber \\
&~& N_q \equiv \frac{i^q}{\Gamma (-q)(1-e^{-i2\pi q})}=
\frac{(-i)^{q+1}\Gamma (q+1)}{2\pi}~,
\end{eqnarray} 
where we have incorporated the velocity coupling $v_i$ in the definition 
of the $\sigma$-model deformation, and 
we have defined the integral along the contour of figure 1, 
having chosen the cut to be from $+i\epsilon$ to $+i\infty$.

\subsection{Extended Logarithmic world-sheet Algebra of recoil in RW 
backgrounds}

Following the flat space time analysis of \cite{kmw}, we now 
proceed to discuss the conformal structure of the 
recoil operators in RW backgrounds. 
We shall do so by acting on the operator ${\cal }D^{(q)}$ (\ref{opd}) with the
world-sheet energy momentum tensor operator $T_{zz} \equiv T$ 
(in a standard notation).  
Due to the form of the background space time (\ref{rwmetric}), 
the stress tensor $T$
assumes the form 
\begin{equation}\label{stress}  
2T =-(\partial t)^2 + a^2(t) (\partial X^i)^2 
\end{equation}
where,  from now on, $\partial \equiv \partial _z$, unless otherwise stated.  
One can then obtain 
the relevant 
operator-product expansions (OPE) of $T$ with 
the operators ${\cal D}^{(q)}$. For convenience in what follows
we shall consider the action of each of the two terms in (\ref{stress})
on the operators ${\cal D}^{(q)}$ separately. For the first 
(time $t$-dependent part), one has, as $z \to w$:
\begin{eqnarray} 
&~& -\frac{1}{2}(\partial t(z))^2\cdot{\cal D}^{(q)}(w) 
=  \frac{v_i}{(z-w)^2}\left[
N_q\int _{-\infty}^{+\infty}d\omega 
\frac{\omega^2/2}{(\omega -i\epsilon)^{q+1}}~e^{i\omega~t(w)}\right]
\partial_nX^i 
= \nonumber \\
&~& \frac{1}{(z-w)^2}\left[-\frac{\epsilon^2}{2}{\cal D}^{(q)} + q\epsilon {\cal D}^{(q-1)} + \frac{q(q-1)}{2}{\cal D}^{(q-2)} \right]
\end{eqnarray} 
The above formul\ae\, were derived for asymptotically large time $t$,
assuming the two-point correlators 
\begin{equation}\label{correl}
\langle X^\mu(z) X^\nu(w) \rangle 
= 2G^{\mu\nu}{\rm ln}|z-w|^2 + \dots~,
\end{equation}
where the $\dots $ denote terms with negative powers of $t$, 
related to space-time curvature,
which are subleading in the limit $t \to \infty$.   

For the spatial part of (\ref{stress}) we consider the OPE 
$a(t(z))^2 (\partial X^i(z))^2 {\cal D}^{(q)}(w)$ as $z \to w$. Again, for 
convenience we shall do the time and space contractions separately:
\begin{eqnarray}
&~& t^{2p}(z)\cdot {\cal D}^{(q)}(w) = \int d\omega {\cal {\tilde D}}^{(q)}(\omega) 
t^{2p}(z)\cdot
e^{i\omega (t(w)-t_0)} = \nonumber \\
&~& \int _0^{\infty} \frac{d\nu}{\Gamma (-2p)} \nu^{-1-2p}
\int d\omega {\cal {\tilde D}}^{(q)}(\omega) e^{-\nu t(z)}\cdot
e^{i\omega (t(w)-t_0)}  
\end{eqnarray} 
Using the OPE $e^{-\nu t(z)} \cdot e^{-i\omega (t(w)-t_0)} \sim 
|z-w|^{i\nu\omega}e^{-\nu t(z) - i\omega (t(z)-t_0) + {\cal O}(z-w)}$
one obtains (as $z \sim w$):
\begin{eqnarray} 
&~& t(z)^{2p}\cdot{\cal D}^{(q)}(w)=
\int _0^\infty \frac{d\nu}{\Gamma (-2p)}\nu^{-1-2p}e^{-\nu t(z)}
{\cal D}^{(q)} (t-t_0 -\nu {\rm ln}|z-w|) = \nonumber \\
&~& t^{2p}~\int _0^\infty \frac{d\nu}{\Gamma (-2p)}\nu^{-1-2p}e^{-\nu}
{\cal D}^{(q)} (t-t_0 -\frac{\nu}{t} {\rm ln}|z-w|) = \nonumber \\ 
&~& t^{2p}\left[{\cal D}^{(q)}(t-t_0) -\frac{1}{t}{\rm ln}|z-w|
\frac{\Gamma (1-2p)}{\Gamma (-2p)}\frac{d}{dt}{\cal D}^{(q)}(t-t_0)
+ {\cal O}(t-t_0)^{q-2}\right]
\end{eqnarray}
We now observe that $\tfrac{d{\cal D}^{(q)}}{dt}=q{\cal D}^{(q-1)} - 
\epsilon~{\cal D}^{(q)} $, where both terms have vacuum expectation
values 
of the same order
in $\epsilon$, as we shall see below, 
and hence both should be kept in our perturbative 
expansion.

Expanding the various terms around $t_0$,  
$t^{s}=(t-t_0)^{s} + s~t_0(t-t_0)^{s-1}+
\frac{t_0^{2}}{2}(s)(s-1)(t-t_0)^{s-2}+{\cal O}([t-t_0]^{s-3})$, 
one has: 
\begin{eqnarray}  
&~& t^{2p}(z)\cdot{\cal D}^{(q)}(w) = {\cal D}^{(2p+q)}(t-t_0)+ 
\left(2p~t_0 - 2p~\epsilon~{\rm ln}|z-w|\right)~{\cal D}^{(2p+q-1)}+ 
\nonumber \\
&~& + \left(\frac{t_0^{2}}{2}2p(2p-1)+[2pq + 
(2p-4p^2)\epsilon~t_0]~{\rm ln}|z-w|\right)
{\cal D}^{(2p+q-2)}(t-t_0)+ \nonumber \\
&~& + {\cal O}([t-t_0]^{2p+q-3})
\end{eqnarray}
where it is worthy of mentioning that inside the subleading terms
there are higher logarithms of the form ${\rm ln}^n|z-w|$, where 
$n =2,3,4, \dots$.

We now come to the OPE between the spatial parts. In view of 
(\ref{correl}), upon expressing $\partial _z$ in normal $\partial_n$ 
and tangential parts,
and imposing Dirichlet boundary conditions on the world-sheet boundary
where the operators live on, we observe that 
such operator products take the form:
\begin{equation}
(\partial X^j(z))^2 \cdot \partial_n X^i(w) \sim 
G^{ii}\frac{1}{(z-w)^2}\partial_n X^i \sim \frac{t^{-2p}}{(z-w)^2}
\partial_n X^i, \qquad ({\rm no}~~{\rm sum}~~{\rm over}~~i)
\end{equation} 
Performing the last contraction with the $t^{-2p}$, following the previous
general formul\ae\, and collecting appropriate terms, one obtains:
\begin{eqnarray}\label{elalg}
&~& T(z)\cdot{\cal D}^{(q)}[(t-t_0)(w)] 
=\frac{1-\frac{\epsilon^2}{2}}{(z-w)^2}{\cal D}^{(q)} [(t-t_0)(w)] +
\frac{q\epsilon}{(z-w)^2}{\cal D}^{q-1}[(t-t_0)(w)] + \nonumber \\
&~& + \frac{\frac{q(q-1)}{2} -2p^{2}~{\rm ln}|z-w| 
-2p^{2}~\epsilon^2~{\rm ln}^2|z-w|}{(z-w)^2}{\cal D}^{(q-2)} 
[(t-t_0)(w)]
+ {\cal O}([t-t_0]^{q-3})
\end{eqnarray}
where again inside the subleading terms there are 
higher logarithms.

We next notice that, 
as a consistency check of the formalism, one can  
calculate the OPE (\ref{elalg}) in case one considers matrix elements
between  
{\it on-shell} physical states. 
In the context of $\sigma$-models, we are working with, the 
physical state condition implies the constraint of the 
vanishing of the world-sheet stress-energy tensor 
$2T=-(\partial t)^2 + a(t)^2(\partial X^i)^2 =0$.
This condition allows $(\partial X^i)^2 $ to be expressed in terms 
of $(\partial t)^2$, which is consistent even at a 
correlation function
level in the case of very target times $t \gg t_0$, since in that case,
the correlator $\langle X^i t \rangle$ 
is subleading, as mentioned previously. 
Implementing this, it can be then seen that the OPE between 
the spatial parts of 
$T$ and ${\cal D}^{(q)}$ is:
\begin{eqnarray} 
&~& a^2(t)(\partial X^i)^2 \cdot {\cal D}^{(q)} = 
t^{-2p}(\partial t)^2 \cdot
\{ {\cal D}^{(2p+q)}(t-t_0)+ 
\left(2p~t_0 - 2p~\epsilon~{\rm ln}|z-w|\right)~{\cal D}^{(2p+q-1)}+ 
\nonumber \\
&~& + \left(\frac{t_0^{2}}{2}2p(2p-1)+[2pq + 
(2p-4p^2)\epsilon~t_0]~{\rm ln}|z-w|\right)
{\cal D}^{(2p+q-2)}(t-t_0)+ \nonumber \\
&~& +{\cal O}([t-t_0]^{2p+q-3})\}.
\end{eqnarray} 
Performing the appropriate contractions, and 
adding to this result 
the OPE of the temporal part of $T$ with ${\cal D}^{(q)}$,
i.e. the quantity 
$-\frac{\epsilon^2}{2}{\cal D}^{(q)} + q\epsilon {\cal D}^{(q-1)} 
+ \frac{1}{2}q(q-1){\cal D}^{(q-2)}$, we obtain:
\begin{eqnarray}
&~& T\cdot {\cal D}^{(q)}|_{\rm on-shell}=\left(-2p\epsilon
-pt_0~\epsilon^2 + p\epsilon^2~{\rm ln}(a/L)\right)~{\cal D}^{(q-1)}
+  \nonumber \\
&~& + \{t_0^2\epsilon^2~2p(2p+1) - 3\epsilon^2p(2p+q){\rm ln}(a/L)
- 2\epsilon^3(p+p^2)t_0{\rm ln}(a/L) - 2p^2\epsilon^4{\rm ln}^2(a/L)
+ \nonumber \\
&~& + \epsilon (2p+q)2p~t_0 - (4p^2 + 4pq - 2p)\}
{\cal D}^{(q-2)} + {\cal O}([t-t_0]^{q-3})
\label{onshell})
\end{eqnarray}
From the above we observe that the on-shell operators become 
marginal as they should, given that an on-shell theory ought to be conformal. 
Moreover, and more important, the world-sheet divergences {\it disappear}
upon imposing the condition 
\begin{equation}\label{xi0}  
\epsilon^{2}{\rm ln}(L/a)^2 = \xi_0 = 
{\rm constant~independent~of}~\epsilon,~a,~L
\end{equation}
where $L$ ($a$) is the world-sheet (ultraviolet) infrared cut-off
on the world sheet.  
As we shall discuss later on, this condition will be 
of importance  
for the closure of the logarithmic algebra, which 
characterizes the fixed point~\cite{kmw}. Hence, 
the conformal invariance is preserved by the on-shell states, any dependence
from it being associated with {\it off-shell} states.

We next notice that, in the context of the RW metric (\ref{rwmetric}), 
there are two cases of expanding universes, 
one corresponding to $0< p \le 1$, and the other
to $p > 1$. 
Whenever $p \le 1$ (which notably incorporates the cases of both 
radiation and matter dominated Universes) 
there is {\it no horizon}, given that 
the latter is given by:
\begin{equation}\label{horizon}
 \delta (t) = a(t)\int_{t_0}^\infty \frac{dt'}{a(t')}
\end{equation} 
In this case the relevant value for $q$ is $q=2p \le 2$.
On the other hand, for the case $p > 1$, i.e. $q > 2$ 
there is a non-trivial cosmological {\it horizon}, which as we shall see 
requires special treatment from a conformal symmetry viewpoint. 

We commence with the no-horizon case, $1 < q \le 2$. 
We first notice that the linear in $t$ term in (\ref{path2}) 
leads to the conventional logarithmic algebra, discussed in
\cite{kmw}, corresponding to a pair of impulse (`recoil') operators
$C,D$. The main point of our discussion below is a study of the 
$t^{2p}$ terms in (\ref{path2}), and their connection to 
other logarithmic algebras. 
Indeed, we observe  that a logarithmic 
algebra~\cite{lcft,kogan,kmw} can be obtained for these 
terms of the operators, if 
we define 
${\cal D} \equiv {\cal D}^{(q)}$ and ${\cal C} \equiv q\epsilon {\cal D}^{(q-1)}$. 
In this case we have the following OPE with $T$:
\begin{eqnarray}\label{tdope}
&~& (z-w)^2~T\cdot {\cal D}=(1-\frac{\epsilon^2}{2}){\cal D} + {\cal C}, \nonumber \\
&~& (z-w)^2~T\cdot {\cal C}=(1-\frac{\epsilon^2}{2}){\cal C} + {\cal O}([t-t_0]^{q-2}),
\end{eqnarray}
where throughout this work 
we ignore terms with negative powers in $t-t_0$ (e.g. of order 
$q-2$ and higher),
for large $t\gg t_0$.  
Notice that in the case $q < 1$ (i.e. $p < 1/2$) the ${\cal C}$ operator 
defined above is absent.

In the second case $p > 1$ one faces the problem of 
having cosmological horizons (cf. (\ref{horizon})), 
which recently has attracted considerable attention in view of 
the impossibility of defining a consistent scattering 
$S$-matrix for asymptotic states~\cite{challenge,emnsmatrix}. 
In this case 
the operator ${\cal D}^{(q-2)}$ is {\it not subleading} 
and one has an {\it extended (higher-order) 
logarithmic algebra} defined by (\ref{elalg}).  
It is interesting to remark that now 
the logarithmic
world-sheet terms in the coefficient of the ${\cal D}^{(q-2)}$ 
operator imply that the limit $z \to w$ 
is plagued by ultraviolet world-sheet divergences, 
and hence the world-sheet conformal invariance 
is spoiled. This necessitates Liouville dressing, in order 
to restore the conformal symmetry~\cite{ddk}. 
As we shall show later, in such cases with horizons 
the recoil of the D-particle 
may induce a non-trivial backreaction on the spacetime geometry, which 
results in an effective spacetime in which the horizons {\it disappear}.
This happens, as we shall discuss later, in the context of Liouville
strings with the identification of the Liouville mode with time.

We now turn to a study of the correlators of the various ${\cal D}^{(q)}$ 
operators, which will complete the study of the associated 
logarithmic algebras, in analogy with the flat target-space case 
of \cite{kmw}.
From the algebra (\ref{elalg}) we observe that we need to evaluate
correlators between ${\cal D}^{(q)}, {\cal D}^{(q-n)},~n=0,1,2, \dots$. 
We shall evaluate correlators $\langle \dots \rangle $ 
with respect to the free world-sheet action,
since we work to leading order in the (weak) coupling $v_i$.
For convenience below we shall restrict ourselves only to the 
time-dependent part of the operators ${\cal D}$. The incorporation 
of the $\partial_n X^i$ is trivial, and will be implied in what follows. 
With these in mind one has: 
\begin{eqnarray}\label{qqn}  
\langle {\cal D}^{(q)}(z) {\cal D}^{(q-n)}(w) \rangle =
N_q N_{q-n} \int \int_{-\infty}^{+\infty}  \frac{d \omega d\omega '}{
(\omega - i\epsilon)^{q+1}(\omega' - i\epsilon)^{q-n+1}}
\langle  e^{-i\omega t(z)}~e^{-i\omega' t(w)} \rangle
\end{eqnarray}  
where $\epsilon \to 0^+$.  
As already mentioned, 
we work to leading order in time $t \gg \infty$, and hence 
we can  
we apply the formula (\ref{correl}) 
for two-point correlators of the $X^\mu$ fields to 
write~\footnote{Here we use simplified propagators on the boundary,
with the latter represented by a straight line; 
this means that the arguments of the logarithms are real~\cite{kmw}.
To be precise, one should use the full expression for the propagator 
on the disc, along the lines of \cite{szabo}. As shown there,
and can be checked here as well, the results are unaffected.}
\begin{eqnarray}
&~& \langle e^{-i\omega t(z)}~e^{-i\omega' t(w)}\rangle  
=e^{-\frac{\omega^2}{2}\langle t(z) t(z)\rangle -
\frac{\omega^{'2}}{2}\langle t(w) t(w)\rangle - 
\omega \omega'\langle t(z) t(w)\rangle} = \nonumber \\
&~& = e^{-(\omega+\omega')^2{\rm ln}(L/a)^2 + 2\omega\omega'{\rm ln}(|z-w|/a)^2}~,
\end{eqnarray} 
where we took into account that ${\rm Lim}_{z\to w}~\langle t(z)t(w)\rangle
= -2{\rm ln}(a/L)^2$. Given that ${\rm ln}(L/a)$ is very large, 
one can approximate 
$e^{-(\omega+\omega')^2{\rm ln}(L/a)^2} \simeq 
\frac{\sqrt{\pi}}{\sqrt{{\rm ln}(L/a)^2}}\delta(\omega+\omega')$.
Thus we obtain:
\begin{eqnarray} 
\langle {\cal D}^{(q)}(z) {\cal D}^{(q-n)}(w) \rangle 
= (-1)^{-q+n-1}~N_q~N_{q-n}{\cal J}_n^{(q)}~,
\qquad {\cal J}_n^{(q)} \equiv \sqrt{\frac{\pi}{\alpha}}\int_{-\infty}^{+\infty}
\frac{d\omega~e^{-\omega^2 
\lambda}~(\omega+i\epsilon)^{n}}{(\omega^2 + \epsilon^2)^{q+1}}
\end{eqnarray}
where $\lambda \equiv 2{\rm ln}(|z-w|/a)^2$, 
and $\alpha \equiv {\rm ln}(L/a)^2$. 

Below, for definiteness, we shall be 
interested in the case $2<q<3$, in which 
the relevant correlators are given by $n=0,1,2$. 
One has: 
\begin{eqnarray}\label{defj} 
&~&{\cal J}_{0}^{(q)}~=\sqrt{\frac{\pi}{\alpha}}~\epsilon^{-2q-1}~f_q (\epsilon^2 \lambda)~;
\nonumber \\
&~& f_q(\xi) =\sqrt{\pi}\frac{\Gamma (\frac{1}{2}+q)}{\Gamma (1 + q)}~F(\frac{1}{2},~\frac{1}{2}-q~;\xi)
+ ~\xi^{\frac{1}{2}+q}~\Gamma(-\frac{1}{2}-q)~F(1+q,~\frac{3}{2}+q~;\xi) 
\nonumber \\
&~&{\cal J}_{1}^{(q)}~=~i\epsilon{\cal J}_{0}^{(q)}~, \nonumber \\
&~& {\cal J}_{2}^{(q)}=-2\epsilon^2~{\cal J}_0^{(q)} + {\cal J}_0^{(q-1)}=-\frac{\partial}{\partial \lambda}{\cal J}_0^{(q)}-\epsilon^2{\cal J}_0^{(q)}
\end{eqnarray} 
where $F(a,b;z)=1 + \frac{a}{b}\frac{z}{1!}+ \frac{a(a+1)}{b(b+1)}\frac{z^2}{2!} + \dots $
is the degenerate (confluent) hypergeometric function. \\
Thus, the form of the algebra away from the fixed point (`{\it 
off-shell form}'), 
i.e. for $\epsilon^2 \ne 0$, is:  
\begin{eqnarray}\label{offshellalg}   
&~& \langle {\cal D}^{(q)}(z) {\cal D}^{(q)}(0) \rangle 
= {\tilde N}_q^2~\sqrt{\frac{\pi}{\xi_0}}~\left(f_q(2\xi_0)
(\frac{\alpha}{\xi_0})^q + 2~f'_q(2\xi_0)(\frac{\alpha}{\xi_0})^{q-1}~
{\rm ln}(|z/L|^2) + \right.\nonumber \\ 
&~&  \left.+ \frac{1}{2}~f''_q(2\xi_0)(\frac{\alpha}{\xi_0})^{q-2}~
4{\rm ln}^2(|z/L|^2)
+ {\cal O}(\alpha^{q-3})\right)~, \nonumber \\
&~&  \epsilon~q~\langle {\cal D}^{(q)}(z) {\cal D}^{(q-1)}(0) \rangle =
{\tilde N}_q^2\sqrt{\frac{\pi}{\xi_0}}~\left(f_q(2\xi_0)
(\frac{\alpha}{\xi_0})^{q-1} + \right. \nonumber \\
&~& \left. + 2~f'_q(2\xi_0)(\frac{\alpha}{\xi_0})^{q-2}~
{\rm ln}(|z/L|^2) + {\cal O}(\alpha^{q-3})\right)~, \nonumber \\
&~&  \epsilon^2~q^2~\langle {\cal D}^{(q-1)}(z) {\cal D}^{(q-1)}(0)\rangle
= {\tilde N}_q^2~\sqrt{\frac{\pi}{\xi_0}}~f_{q-1}(2\xi_0)
(\frac{\alpha}{\xi_0})^{q-2} + {\cal O}(\alpha^{q-3})
, \nonumber \\
&~&  \epsilon^2~q~(q-1)~\langle {\cal D}^{(q)}(z) {\cal D}^{(q-2)}(0)\rangle
= -{\tilde N}_q^2~\sqrt{\frac{\pi}{\xi_0}}  
\left(f_q(2\xi_0) + f'_q(2\xi_0)\right)~(\frac{\alpha}{\xi_0})^{q-2}
+ {\cal O}(\alpha^{q-3})~, \nonumber \\
&~& \epsilon^3~q^2(q-1)~\langle {\cal D}^{(q-1)}(z) {\cal D}^{(q-2)}(0)\rangle
= {\cal O}(\alpha^{q-3})~, 
\nonumber \\
&~& \epsilon^4~q^2(q-1)^2~\langle {\cal D}^{(q-2)}(z) {\cal D}^{(q-2)}(0)\rangle
= {\cal O}(\alpha^{q-4})~
\end{eqnarray} 
where ${\tilde N}_q = \frac{\Gamma (1+q)}{2\pi}$, and 
$\xi_0$ has been defined in (\ref{xi0}). 

Notice that the above algebra is plagued by world-sheet 
ultraviolet divergences as $\epsilon^2 \to 0^+$, thereby 
making the approach to the fixed (conformal) point subtle. 
As becomes obvious from (\ref{xi0}), the non-trivial fixed point 
$\epsilon \to 0^+$ 
corresponds to $L/a \to +\infty$, i.e. it is an infrared world-sheet
fixed point. 
In order to understand the approach to the infrared fixed point,
it is important to make a few remarks first,
motivated by physical considerations. 

From the integral expression of the regularized 
Heaviside function~\cite{kmw} 
(\ref{rep}) it becomes obvious that 
a scale $1/\epsilon$ for the target time
is introduced. This, together with the fact that 
the scale $\epsilon$ is connected (\ref{xi0}) to the 
world renormalization-group scales $L/a$, implies naturally
the introduction of a `renormalized' $\sigma$-model coupling/velocity
$v_{R,i}(\tfrac{1}{\epsilon}) $ at the scale $\tfrac{1}{\epsilon}$:
\begin{equation} 
v_{R,i}(\frac{1}{\epsilon}) \sim \left(\frac{1}{\epsilon}\right)^{q-1}
\label{velrenorm}
\end{equation}
for a trajectory $y_i (t) \sim t^q$. 
This normalization 
would imply the following 
rescaling of the operators 
\begin{equation}
{\cal D}^{(q-n)} \rightarrow \epsilon^{q-1}{\cal D}^{(q-n)}
\end{equation} 
As a consequence, the 
factors $\epsilon^{2(1-q)}$ in (\ref{defj}), (\ref{offshellalg})  
are removed.  
In the context of the world-sheet field theory this renormalization 
can be interpreted as a subtraction of the ultraviolet 
divergences by the addition of appropriate counterterms
in the $\sigma$ model.

The approach to the infrared fixed point $\epsilon \to 0^+$ 
can now be made by looking at the {\it connected} 
two point correlators between the operators ${\cal D}^{(q)}$ defined by
\begin{equation}\label{connected}  
\langle {\cal A}{\cal B} \rangle_c = \langle {\cal A}{\cal B} \rangle - 
\langle {\cal A} \rangle~\langle {\cal B} \rangle~,
\end{equation}
where the one-point functions are given by:
\begin{eqnarray} 
&~&\langle {\cal D}^{(s)}\rangle = N_s\int 
\frac{d\omega}{(\omega -i\epsilon)^{s+1}}\langle e^{i\omega t} \rangle 
=N_s\int \frac{d\omega}{(\omega -i\epsilon)^{s+1}} e^{-\omega^2\alpha} = {\tilde N}_s~\epsilon^{-s}~h_s(\epsilon^2\alpha)~, \nonumber \\
&~&h_s(x) 
= -\frac{x^{s/2}}{2}\left(\frac{4\pi}{\Gamma (\frac{1+s}{2})}~
\sqrt{\pi}~F(1+\frac{s}{2}~, \frac{3}{2}~, x) - \frac{2\pi}{\Gamma (1 + 
\frac{s}{2})}~F(\frac{1 + s}{2}~, \frac{1}{2}~, x)\right).
\end{eqnarray}  
 
For the two-point function of the ${\cal D}^{(q)}$ operator 
the result is: 
\begin{eqnarray} 
&~& \langle {\cal D}^{(q)}(z){\cal D}^{(q)}(0) \rangle_c =  
{\tilde N}_q \epsilon^{-2}\left(\frac{\sqrt{\pi}}{\xi_0}
f_q(2\xi_0+2\epsilon^2{\rm ln}|z/L|^2) - h^2_q(\xi_0)\right)~.
\end{eqnarray}

Expanding in powers of $\epsilon$, we obtain  
\begin{eqnarray} \label{dd}
&~& \langle {\cal D}^{(q)}(z) {\cal D}^{(q)}(0) \rangle_c =  
{\tilde N}_q \epsilon^{-2}\left(\frac{\sqrt{\pi}}{\sqrt{\xi_0}}
f_q(2\xi_0) - h^2_q(\xi_0)\right) + 
{\tilde N}^2_q~\frac{\sqrt{\pi}}{\sqrt{\xi_0}}~f'_q(2\xi_0)2
{\rm ln}|z/L|^2 + \nonumber \\
&~& +~ \epsilon^2 {\tilde N}_q^2 \sqrt{\frac{\pi}{\xi_0}}
\frac{1}{2}f''_q(2\xi_0)4 {\rm ln}^2|z/L|^2
+ \dots  
\end{eqnarray} 
where $\dots $ denote terms that vanish as $\epsilon \to 0^+$.

To avoid the divergences coming from the $\epsilon^{-2}$ factors,
the following condition must be satisfied: there must be a 
solution $\xi_0=\xi_0(q)$ of the equation: 
${\cal H}(\xi_0) \equiv \frac{\sqrt{\pi}}{\sqrt{\xi_0}}
f_q(2\xi_0) - h^2_q(\xi_0)=0$. The existence of such a solution 
can be verified numerically 
(see figure 2). Analytically this can be confirmed by 
looking at the asymptotic behaviour of the function 
${\cal H}(x)$ as $x \to \infty$, which yields a negative value: 
${\cal H}(x\to \infty) \sim 
-\frac{\pi^3x^{2q}e^{2x}}{\Gamma^2 (\frac{1+q}{2})  
\Gamma^2(1 + \frac{q}{2})} < 0$. This behaviour comes entirely from 
the term $h_q^2(x)$, given that $f_q(x\to \infty) \to 0^+$. 

\begin{figure}[t] 
\begin{centering} 
\epsfig{figure=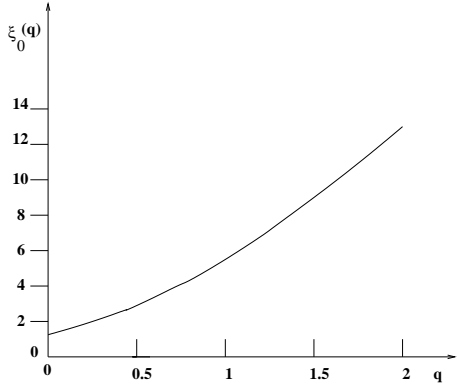, width=0.4\linewidth}
\caption{{\it Graphic solution 
of the equation $\frac{\sqrt{\pi}}{\sqrt{\xi_0}}
f_q(2\xi_0) - h^2_q(\xi_0)=0$.}}
\end{centering} 
\label{sol}
\end{figure}

As we shall show below, for various values of $q$,
near the fixed point $\epsilon \to 0^+$, 
one can construct higher order logarithmic algebras, 
whose highest power is determined by the dominant 
terms in the operator algebra of correlators (\ref{offshellalg}),
(\ref{tdope}). 
To this end, we first remark that in the above analysis
we have dealt with a small but otherwise arbitrary 
parameter $\epsilon$, which allows us to keep 
as many powers as required by (\ref{offshellalg})
in conjunction with the value of $q$. The value of $\epsilon$ 
determines the distance from the fixed point.

For $1< q <2$, there are only two dominant 
operators as the time $t \to \infty$, ${\cal D}$,${\cal C}$.   
In this case 
one 
obtains a conventional logarithmic conformal algebra
of two-point functions
near the fixed point: 
\begin{eqnarray}\label{logalg}
&~& \langle {\cal D}^{(q)}(z)~{\cal D}(0)^{(q)} 
\rangle_c = \langle {\cal D}(z)~{\cal D}(0) \rangle_c  
\sim {\tilde N}^2_q~\frac{\sqrt{\pi}}{\sqrt{\xi_0}}~f'_q(2\xi_0)2
{\rm ln}|z/L|^2~, \nonumber \\
&~& \epsilon~q~\langle {\cal D}^{(q-1)}(z)~D^{(q)}(0) \rangle_c = 
\langle {\cal C}(z)~{\cal D}(0) \rangle_c \sim {\tilde N}^2_q~
\left(h^2_q(\xi_0) - h_{q-1}~h_q (\xi_0)\right)~, 
\end{eqnarray} 
and all the other correlators are subleading as $t \to \infty$.

Therefore, the {\it on shell algebra} is of the conventional 
{\it logarithmic form}~\cite{lcft}, between a pair of operators,  
and hence,
${\cal D}^{(q-2)}$ and subsequent
operators, which owe their existence to the non-trivial RW metric,
do not modify the two-point correlators of the standard
logarithmic algebra of `recoil' 
(impulse)~\cite{kmw}~\footnote{We note at this stage that, in our case of non-trivial cosmological 
RW spacetimes, the pairs of operators ${\cal D},{\cal C}$ do not represent
velocity and position as in the flat space time case of ref. \cite{kmw},
but rather velocity and acceleration. 
This implies that, under a finite-size scaling of the world sheet,
the induced transformations of these operators do not form 
a representation of the Galilean transformations of the flat-space-time 
case.}. 

Next, we consider the case where $2< q < 3$. In this case, from 
(\ref{offshellalg}) we observe that there are now three operators
which dominate in the limit $t \to \infty$,  
${\cal D}$, ${\cal C}$ and ${\cal B} = \epsilon^2~q~(q-1)
{\cal D}^{(q-2)}$, whose form is implied from (\ref{tdope}), in 
analogy with ${\cal C}$. The corresponding algebra of
correlators consists of parts forming 
a conventional logarithmic algebra, and parts 
forming a second-order logarithmic algebra, the latter being 
obtained
from terms of order $\epsilon^2$ in the appropriate
two-point connected correlators ( cf. (\ref{dd}) {\it etc.}),
which are denoted by a superscript $\langle \dots \rangle_c^{(2)}$:
\begin{eqnarray}\label{higherord}  
&~& \langle {\cal D}(z)~{\cal D}(0) \rangle_c^{(2)} = 
{\tilde N}_q^2 \sqrt{\frac{\pi}{\xi_0}}
\frac{1}{2}f''_q(2\xi_0)4 {\rm ln}^2|z/L|^2~, 
\nonumber \\
&~& \langle {\cal C}(z)~{\cal D}(0) \rangle_c^{(2)}= 
{\tilde N}_q^2 \sqrt{\frac{\pi}{\xi_0}}
2f'_q(2\xi_0) {\rm ln}|z/L|^2~, 
\nonumber \\
&~& \langle {\cal C}(z)~{\cal C}(0) \rangle_c^{(2)} = 
{\tilde N}_q^2 \sqrt{\frac{\pi}{\xi_0}}
f_{q-1}(2\xi_0)~, \nonumber \\
&~& \langle {\cal B}(z)~{\cal D}(0) \rangle_c^{(2)} = 
-{\tilde N}_q^2 \sqrt{\frac{\pi}{\xi_0}}
\left(f_{q}(2\xi_0) + f'_q(2\xi_0)\right)~, \nonumber \\
&~& \langle {\cal C}(z)~{\cal B}(0) \rangle_c^{(2)} = 
\langle {\cal B}(z)~{\cal B}(0) \rangle_c^{(2)} = 0
\end{eqnarray} 
where the last two correlators are of order $\epsilon^4$ and $\epsilon^6$
respectively, that is of higher order than the $\epsilon^2$ terms,
and hence they are viewed as zero to the order we are working here.

In general, if one considers $q > 3$ one arrives at higher order
logarithmic algebras~\cite{lcft}, 
with the highest power given by the 
integer value of $q$, $[q]$. This is an interesting 
feature 
of the recoil-induced  
motion of $D$-particles in RW backgrounds with scale
factors $\sim t^{p}$, $p >1$, corresponding to cosmological horizons
and accelerating Universes. In such a case the order of 
the logarithmic algebra is given by $[2p]$.    
It is interesting to remark that radiation and matter (dust) dominated 
RW Universes would imply simple logarithmic algebras.
 
We now notice that,  
under 
a world-sheet finite-size scaling,  
\begin{eqnarray} 
L \to L' = L~e^{{\cal T}\,{\cal K}(q)}~, \qquad \epsilon^{-2} 
\to (\epsilon')^{-2} = \epsilon^{-2} + {\cal T}   
\end{eqnarray} 
with ${\cal K}(q)$ a function of $q$ determined by 
(\ref{logalg}), 
the operators ${\cal C},~{\cal D}, \dots $, and consequently the 
target-time $t$, 
transform in a non trivial way. In particular, for $t$ one has: 
\begin{eqnarray}\label{timeshift2}  
\left(\frac{\epsilon'}{\epsilon}\right)^{q-1}{\cal Z}~({\cal T})^q 
t({\cal T})^q = t^q + q~\epsilon~{\cal T}~t^{q-1} + {\cal O}(\epsilon^2)
\end{eqnarray}
where ${\cal Z}({\cal T})$ 
is a wave function renormalization of the world-sheet 
field $t(z)$, which can be chosen in a natural way so that 
$\left(\frac{\epsilon'}{\epsilon}\right)^{q-1}{\cal Z}\,({\cal T})^q =1$. 
This implies 
\begin{eqnarray}\label{timeshift}  
&~& t({\cal T})^q = (t + \epsilon {\cal T})^q + {\cal O}(\epsilon^2)~,
\nonumber \\
&~& t({\cal T}) = t + \epsilon {\cal T} + {\cal O}(\epsilon^2)~,
\end{eqnarray}
i.e. that a shift in the target time is represented as $\epsilon\,{\cal T}$.
Of course, at the fixed point, $\epsilon =0$, 
the field $t(z)$ does not run, as expected. 
As we shall see in the next section, 
the shift (\ref{timeshift}) is consistent 
with the identification of the Liouville mode with the target time, 
in case one wishes to discuss certain aspects of slightly 
off-shell string physics. 

\section{Space Time Metrics}

\subsection{Vertex Operator for the Path and associated SpaceTime Geometry}

In this section we shall discuss the implications of the world-sheet
deformation (\ref{path}) for the spacetime geometry. In particular, 
we shall show that its r\^ole is to preserve the Dirichlet boundary 
conditions on the $X^i$ by changing coordinate system, which is 
encoded in an induced change in the space time geometry $G_{ij}$. 
The final coordinates, then, are coordinates in the rest frame 
of the recoiling particle, which naturally explains the 
preservation of the Dirichlet boundary condition.  

To this end, we first rewrite 
the world-sheet boundary vertex operator (\ref{path}) as
a bulk operator: 
\begin{eqnarray} \label{bulkop}
&~& V = \int _{\partial \Sigma} G_{ij}y^j(t)\partial_n X^i =
\int_{\Sigma} \partial_\alpha \left(y_i(t)\partial^\alpha X^i\right) =
\nonumber \\
&~& = \int _{\Sigma} \left({\dot y}_i(t)\partial_\alpha t \partial^\alpha X^i 
+ y_i \partial^2 X^i \right)
\end{eqnarray}
where the dot denotes derivative with respect to the target time $t$, and 
$\alpha$ is a world-sheet index. Notice that it is the covariant 
vector $y_i$ which appears in the formula,
which incorporates the metric $G_{ij}$, $y_i=G_{ij}y^j$. 

To determine the 
background geometry, which the string is moving in, it is sufficient
to use the classical motion of the string, described by the 
world-sheet equations: 
\begin{eqnarray}\label{sem} 
\partial^2 X^i + {\Gamma^i}_{\mu\nu}\partial_\alpha X^\mu 
\partial^\alpha X^\nu =0,
\end{eqnarray}
where $\mu, \nu$ are space time indices, $\alpha=1,2$ is a world-sheet
index, $\partial^2$ is the Laplacian on the world sheet, 
and $i$ is a target spatial index.

The relevant Christoffel symbol in our RW background case, is 
${\Gamma^i}_{ti}$, and thus the operator (\ref{bulkop}) becomes: 
\begin{eqnarray} 
\int _{\Sigma} \left({\dot y}_i - 2y_i(t){\Gamma^i}_{ti}\right)\partial_\alpha t \partial^\alpha X^i 
\end{eqnarray} 
from which we read an induced non-diagonal component for the space time metric 
\begin{eqnarray}\label{indmetr} 
2G_{0i}= {\dot y}_i - 2y_i(t){\Gamma^i}_{ti}
\end{eqnarray}
In the RW background (\ref{rwmetric}) the path $y_i(t)$  
is described (\ref{pathexpre}) by 
(notice again we work with covariant vector $y_i$):
\begin{eqnarray} 
y_i(t) = \frac{v_ia_0^2}{1-2p}\left(t t_0^{2p}-t_0 t^{2p}\right)
\end{eqnarray}
which gives $2G_{0i}=a^2(t_0) v_i$, yielding for the metric line element:
\begin{equation}\label{fixedpoint}
ds^2=-dt^2 + v_ia^2(t_0)dtdX^i + a^2(t)(dX^i)^2, \qquad {\rm for} 
\qquad t > t_0
\end{equation} 
As expected, this spacetime has precisely the form 
corresponding to a Galilean-boosted frame (the D-particle's rest frame), 
with the boost occurring suddenly at time $t=t_0$.  

This can be understood in a general fashion by first noting that 
(\ref{indmetr}) can be written in a general covariant 
form as:
\begin{eqnarray}
2G_{0i}=\nabla_t y_i~~(= \nabla_t y_i + \nabla_i t)
\end{eqnarray} 
which is the general coordinate transformation associated with $y_i$ 
from a passive (Lie derivative) point of view. 

In general, given the boundary condition 
$\partial_n t=0$, one can 
write the operator (\ref{path}), 
in a covariant form by expressing it as a world-sheet bulk operator:
\begin{eqnarray}
V= \int _{\partial \Sigma} y_\mu \partial_n X^\mu =
\int _{\Sigma} \partial_\alpha \left(y_\mu \partial^\alpha X^\mu \right) = \int_{\Sigma} \nabla_{\mu} y_\nu 
\partial_\alpha X^\mu \partial^\alpha X^\nu  
\end{eqnarray}
where in the last step, we have used again the string equations of 
motion (\ref{sem}). From this expression, one then derives the 
induced change in the metric
\begin{eqnarray}\label{lie} 
2 \delta G_{\mu\nu} = \nabla_\mu y_\nu + \nabla_\nu y_\mu  
\end{eqnarray}  
which is the familiar expression of the Lie derivative under the coordinate 
transformation associated with $y_\mu$. 

In all the above expressions we have taken the limit $\epsilon \to 0$,
which corresponds to considering the ratio 
of world-sheet cut-offs $a/L \to 0$,
implying that one approaches the infrared fixed point in a Wilsonian sense.
As noted previously, in the context of the logarithmic 
conformal analysis of the path $y^i(t)$, we have seen that this limit 
can be reached without problems only in the case $p \le 1$,  which 
corresponds to the absence of cosmological horizons. 
On the other hand,
the case of non-trivial horizons, $p > 1$,
implies 
ultraviolet divergences, which prevent 
one from taking this limit in a way consistent with conformal 
invariance of the  underlying $\sigma$ model. In such a case, the 
operators are relevant, with finite anomalous dimensions $-\epsilon^2/2$,
and thus Liouville dressing is required~\cite{ddk,recoil}. 
This is the topic of the next subsection. 

\subsection{Cosmological Horizons and Liouville Dressing} 

In this subsection we shall discuss Liouville 
dressing of the relevant recoil deformations~\cite{recoil}. 
There are two 
ways one can proceed in this matter. The first, concerns
dressing of the boundary operators (\ref{path}) 
\begin{eqnarray} \label{boundary}
V_{L,{\rm boundary}} = \int _{\partial \Sigma} e^{\alpha_i \varphi} y_i(t)\partial_nX^i, 
\qquad  \alpha_i = -\frac{Q}{2} + \sqrt{\frac{Q^2}{2} + (1-h_i)} 
\end{eqnarray} 
where $h_i$ is the boundary conformal dimension, and
$Q^2$ is the induced central charge deficit on the boundary 
of the world-sheet. 

In a similar spirit to the flat target-space case~\cite{kmw}, 
the rate of change of 
$Q^2$ with respect to world-sheet scale ${\cal T} \sim \epsilon^{-2}$ 
is given by means of Zamolodchikov C-theorem~\cite{zam},
and it is found to be of order~\cite{recoil} $v_i^2~\epsilon^4 $,
as being proportional to the square of the renormalization-group 
$\beta^i$ functions ($i = v_i$): 
$\tfrac{\partial Q^2}{\partial {\cal T}} \propto
-\beta^i {\cal G}_{ij} \beta^j $, where ${\cal G}_{ij} = \delta _{ij} + \dots$,
is the Zamolodchikov metric in coupling constant space.
This implies that $Q^2(t) = Q_0^2 + {\cal O}(\epsilon^2)$,
where $Q_0^2$ is constant. 

We shall distinguish two cases for $Q_0$. The first concerns the case
where $Q_0 \ne 0$ (and by appropriate normalization may be assumed to be 
of order ${\cal O}(1)$). This is the case 
of strings living in a non-critical space time dimension.  
The other pertains to the case where
the only source of non-criticality is the impulse deformation,
i.e. $Q_0 =0$. 
In the former case, 
one has a Liouville dimension $\alpha_i \sim \epsilon^2 $,
while in the latter $\alpha_i \sim \epsilon$. 
In {\it both cases} however, $\epsilon \sim \tfrac{1}{t}$,
where $t$ is the target time.

In the second method~\cite{recoil}, one dresses by the Liouville field 
the 
bulk operator (\ref{bulkop}), i.e. 
\begin{eqnarray} \label{bulk} 
V_{L, {\rm bulk}} = \int _{\Sigma} e^{\alpha_i \varphi} 
\partial_\alpha \left(y_i(t)\partial^\alpha X^i \right), \qquad 
\alpha_i = -\frac{Q}{2} + \sqrt{\frac{Q^2}{2} + (2-\Delta_i)}
\end{eqnarray} 
where $\Delta_i$ is the conformal dimension of the bulk operator.
The central charge deficit $Q$ is of the same order 
$Q^2 = Q_0^2 + {\cal O}(\epsilon^2)$  
as in the boundary case,  which implies again that
$\alpha_i \sim \epsilon^2$ if $Q_0 \ne 0$, and  
$\alpha_i \sim \epsilon$ if $Q_0 = 0$.
An interesting question, which we shall answer in the affirmative 
below, concerns the equivalence
between these two approaches either at the fixed point ($\epsilon \to 0$),
or close to it ($\epsilon \ne 0$ but small). 
 
We commence our analysis by first looking at the boundary operator 
(\ref{boundary}). We may rewrite it as a bulk operator and then manipulate it 
as follows: 
\begin{eqnarray}\label{boundary2} 
&~& V_{L, {\rm boundary}} = \int _{\Sigma} \partial_\alpha 
\left(e^{\alpha_i\varphi}y_i(t)\partial^\alpha X^i\right)=
\nonumber \\
&~& = \int_\Sigma \alpha_i e^{\alpha_i\varphi}~y_i(t)\partial_\alpha \varphi
\partial^\alpha X^i + 
\int_\Sigma e^{\alpha_i\varphi} {\dot y}_i(t) \partial_\alpha 
t \partial^\alpha X^i + \int _\Sigma e^{\alpha_i\varphi}y_i(t)\partial^2 X^i 
\end{eqnarray} 
For the bulk operator (\ref{bulk}) one has:
\begin{eqnarray} \label{bulk2} 
&~& V_{L, {\rm bulk}} = \int _{\Sigma} 
\partial_\alpha 
\left(e^{\alpha_i\varphi}y_i(t)\partial^\alpha X^i\right) - 
 \int_\Sigma \alpha_i e^{\alpha_i\varphi}~y_i(t)\partial_\alpha \varphi
\partial^\alpha X^i = \nonumber \\
&~& =\int_{\partial \Sigma} e^{\alpha_i\varphi} y_i(t) \partial_n X^i
- \int_\Sigma \alpha_i e^{\alpha_i\varphi}~y_i(t)\partial_\alpha \varphi
\partial^\alpha X^i
\end{eqnarray}  

The logarithmic algebra, as discussed in \cite{kmw} and above, 
implies a 
non-trivial infrared fixed point, which in the case $Q_0 \ne 0$ 
is determined by  
$\varphi_0 = \epsilon^{-2} \sim {\rm ln}(L/a)^2 \to \infty$,
where $\varphi_0$ is the Liouville field world-sheet zero mode.
Thus, $\alpha_i \varphi_0 $ is finite as $\epsilon \to 0^+$. 
Therefore, as expected from the restoration 
of the conformal invariance by means of the Liouville
dressing, one can now take  
safely the infra-red limit $\epsilon \to 0^+$ 
in the above expressions. It is then easy to see that 
one is left {\it in both cases} with the metric  (\ref{fixedpoint}),
thereby proving the equivalence of both approaches
at the infrared fixed point. 

In the case $Q_0 =0$, the running central charge deficit 
$Q^2 ={\cal O}(\epsilon^2)$. Recalling~\cite{ddk} that the above 
formul\ae\, imply a rescaling of the Liouville mode
by $Q \sim \epsilon$, so as to have a canonical kinetic $\sigma$-model
term~\footnote{Notice that this rescaling becomes a trivial 
one in the case where $Q_0 \ne 0$.}, and that in this case it is 
the $\varphi_0/Q$ which is identified with ${\rm ln}(L/a)^2 \sim \epsilon^{-2}$ as pertaining to the covariant world-sheet cutoff, 
one observes that again
$\alpha_i~\varphi$ is finite as $\epsilon \to 0^+$, and hence
similar conclusions are reached concerning the 
equivalence of the two methods of Liouville dressing of the 
impulse operator.   

This equivalence is also valid {\it close to}, but not exactly at, the
infrared 
fixed point, as we demonstrate now. 
To this end, 
we discuss the two cases $Q_0 =0$ and $Q_0 \ne 0$
separately.

Consider first the case $Q_0 \ne 0$. 
In this case $\alpha_i \sim \epsilon^2$, 
$\varphi_0 \sim \epsilon^{-2}$ and hence 
$\alpha_i \varphi_0 \sim \epsilon t = {\rm const}$.
We identify now the Liouville direction $\varphi$ 
with that of the target 
time~\cite{emn,recoil}. Given that $t \sim 
\tfrac{1}{\epsilon}$ this implies that 
$\varphi \sim t^2$.    
Under this identification we observe~\cite{recoil}
in both cases (\ref{boundary2}), (\ref{bulk2}) that 
the terms $e^{\alpha\varphi_0}$, and  
the exponential 
factors $e^{-\epsilon~(t-t_0)}$ appearing in the regulated
$\Theta$ functions~\cite{kmw}
are all of order one. 

From these considerations one obtains an induced non-diagonal metric element 
$G_{0i}$, which in the case (\ref{boundary2}) is
\begin{eqnarray}\label{ndmetric} 
G_{0i}d\varphi dX^i =
\left(v_i~\alpha_i t^{2p} + v_ia^2(t_0)\right)~d(t^2)dX^i 
\simeq v_i~\alpha_i t^{2p}
d(t^2)dX^i~, 
\qquad t \gg t_0
\end{eqnarray} 
and in the case (\ref{bulk2}): 
\begin{eqnarray} 
G_{0i}d\varphi dX^i \simeq -v_i~\alpha_i t^{2p}d(t^2)dX^i~, 
\qquad t \gg t_0
\end{eqnarray} 
We then observe that, up to an irrelevant sign, the 
two results are equivalent in the regime of large $t$, where
our perturbative string (world-sheet) analysis is valid, thereby 
proving the equivalence of the two ways of Liouville dressing
even away from the fixed point.  

Under the fact that one 
identifies $\epsilon^{-1} =t-t_0 \sim t \gg t_0$, 
the non-diagonal element of the spacetime metric becomes:
\begin{eqnarray}\label{ndmetric2}
G_{0i} \sim v_i t^{2p-1} 
\end{eqnarray} 
We remind the reader that we analyze here the case with horizon, 
which implies $p > 1$.  

It is convenient now to diagonalize the metric, which implies
the following line element
\begin{eqnarray} \label{nontrans}
ds^2 = -\frac{v_i^2}{a_0^2}t^{2p-2}dt^2 + 
a_0^2t^{2p}(dX^i)^2 
\end{eqnarray} 
By redefining the time coordinate to $t'=\frac{{v}_i}{a_0 p}t^p$ 
one obtains the induced line element:
\begin{eqnarray}\label{rwfinal} 
ds^2 = -(dt')^2 + \frac{a_0^4~p^2}{{v}_i^2}(t')^2~(dX^i)^2, \qquad t \gg t_0  
\end{eqnarray} 
From (\ref{horizon}), we thus observe that the induced metric  
has {\it no horizon}, and no cosmic acceleration. 
In other words a recoiling $D$-particle, 
embedded in a space time which initially appeared to have an horizon,
back reacted in such a way so as to remove it! Equivalently,
we may say that  
recoiling $D$-particles are consistent only in spacetimes 
without cosmological horizons. 

Similar conclusions are reached in the case $Q_0 =0$. 
In that case, $\varphi/Q \sim \epsilon^{-2}$, as explained above,  
and since $Q \sim \epsilon \sim \tfrac{1}{t}$, one now 
has that $\varphi \sim t$. Again, the exponential terms 
$e^{\alpha_i \varphi}$, $\alpha_i \sim \epsilon$, 
and those coming from the 
regulated $\Theta_\epsilon (t)$ 
are of order one. 
Evidently, the induced non-diagonal
metric has the same form (\ref{ndmetric2}) 
as in the case with $Q_0 \ne 0$,
and one can thus repeat the previous analysis, implying removal 
of the cosmological horizon and stopping of cosmic acceleration.

The reader must have noticed that the same conclusion is 
reached already at the level of the metric (\ref{nontrans}),
before the time transformation, once one interprets the coefficient 
of the $(dt)^2$ as a time-dependent light velocity.
The fact that such situations arise `suddenly', after a time moment $t_0$, 
might prompt the reader to draw some analogy 
with the scenaria of time-dependent 
light velocity, involving some sort of phase transitions 
at a certain moment in the (past) history of our Universe~\cite{moffat}.
In our case, as we have seen,  
one can perform (at late times) 
a change in the time coordinate in order to arrive at a RW metric 
(\ref{rwfinal})~\footnote{There is a slight point to which 
we would like to 
draw the reader's attention. This regards the fact that such transformations 
depend on the recoil velocity, and thus on the energy content
of the matter incident on the D-particle. In case one has a 
`foam' situation~\cite{recoil}, in which  
several incident 
particles interact with collections of $D$-particles,
which are virtual quantum excitations of the string/brane vacuum, 
it is unclear how the present results are modified, and hence it might be 
that one cannot perform simultaneous transformations to diagonalize the metric,
thereby obtaining non-trivial refractive indices~\cite{sarkar}.
Such issues fall beyond the scope of the present article.}. 

The removal of the horizon would seem to imply 
from a field-theoretic point of view
that one can define asymptotic states and thus a proper $S$-matrix.  
However, in the context of Liouville strings, with the Liouville mode 
identified
with the time~\cite{emn}, 
there is no proper $S$-matrix, independently of the existence 
of horizons~\cite{emnsmatrix}. This has to do with the structure
of the correlation functions of vertex operators in this construction,
which are defined over steepest-descent 
closed time-like paths in a path-integral formalism,
resembling closed-time paths of non-equilibrium 
field theory~\cite{emn,emnsmatrix}. In such constructions one can define
properly only a (non factorizable) 
superscattering matrix \$$\ne S\,S^\dagger$.

\section{Conclusions}

In this work we have analyzed in some detail the 
problem of impulse(recoil)-induced motion of a heavy $D$-particle
in a Robertson-Walker spacetime, at large times $t$ after
the moment of impact. We have shown, that for 
RW spacetimes with scale factors $\sim t^p$, there is an order 
$[2p]$-logarithmic algebra, involving a group of 
impulse operators, which are relevant from a world-sheet
renormalization group point of view. 

A detailed study of how one can approach the non-trivial infrared
fixed point is given. In the case where $p > 1$, which is 
physically characterized by the presence of 
cosmological horizons, one encounters world-sheet 
divergences. A proper subtraction of such divergences
is subtle, and a detailed discussion of how this can be done 
has been presented. The fact that away from the fixed point 
the deformed theory is plagued by relevant deformations, 
of anomalous dimension which itself depends on the world-sheet 
renormalization-group scale, implies the need for 
Liouville dressing. 

Such a dressing results in the interesting possibility of 
identifying the Liouville mode with the target time, 
in which case one has a formal description -in terms of conformal
field theory methods on the world sheet - of back reaction effects
of the recoiling $D$-particle on the surrounding space time.
It is interesting to notice that the effect is equivalent to 
a `phase transition' at the moment of impact, in which 
there is induced a time-varying speed of light, effectively 
leading to the removal of the initial cosmological horizon,
and the eventual stopping of the acceleration of the Universe. 

From a field-theoretic view point, 
this would imply 
that in such models proper asymptotic states, and thus an $S$-matrix, 
could be defined. However, from our stringy point of view,
the definition of an $S$-matrix
is still a complicated issue, since the underlying theory is of 
Liouville (non-equilibrium) type~\cite{emnsmatrix}. 
Whether such toy models are of relevance to realistic 
stringy cosmologies remains to be seen.
Nevertheless, we believe that the results presented here, although 
preliminary, are of sufficient interest to prompt further studies
along the directions suggested in this work.

\section*{Acknowledgements} 

We thank R. Szabo for discussions. 
The work of E.G. is supported 
by a King's College London Research Studentship (KRS).

\end{document}